\documentclass[twocolumn,showpacs,preprintnumbers,amsmath,amssymb]{revtex4}

\pdfoutput=1

\usepackage{color}
\usepackage{graphicx}
\usepackage{dcolumn}
\usepackage{bm}
\usepackage{wrapfig}
\usepackage{graphicx}

\newcommand{\ljump} {[ \hspace{-0.08 cm} | }
\newcommand{\rjump}{ |\hspace{-0.08 cm}]}
\newcommand{\be}{\begin{equation}}
\newcommand{\ee}{\end{equation}}
\newcommand{\ba}{\begin{array}}
\newcommand{\ea}{\end{array}}
\newcommand{\vk}{\vspace {.3cm}}

\newcommand{\vks}{\vspace {.1cm}}

\begin{document}


\title{Cohesion-decohesion asymmetry in geckos}

\author{G. Puglisi$^1$ and L. Truskinovsky$^2$}
\affiliation{
$^{1}$Dipartimento di Scienze dell'Ingegneria Civile e dell'Architettura (ICAR), Politecnico
di Bari, Italy}

\affiliation{
$^{2}$LMS, Ecole Polytechnique, 91128, Palaiseau, France}

\begin{abstract}
\noindent Lizards and insects can strongly attach to walls
and then detach applying negligible additional forces. We propose a
simple mechanical model of this phenomenon which implies active  muscle
control.  We show that the detachment force may depend not only on the  properties of the adhesive units, but also on the elastic interaction among these units. By regulating the scale of such cooperative interaction, the organism can actively switch between two modes of adhesion: delocalized  (pull off) and localized (peeling).

\end{abstract}

\maketitle

\section{Introduction}
Active mechanisms involved in biological adhesion
in living systems are of broad theoretical interest in view of potential applications in bio-inspired adhesion devices.
 One of the most challenging issues concerns the reconciliation of strong adhesion  \cite{SN} with easy detachment \cite{TP,CW}.

 Experimental studies reveal that  biological adhesion  at the organismic level is typically mediated by fibrillar microstructures which ensure a molecular level contact in the presence of surface roughness. The necessity of avoiding clustering entails a hierarchy of adhesion devices spanning a wide range of scales \cite{Per2, PBA}. In the  case of geckos the macroscopic adhesion force at the level of a foot results from smaller forces at the scale of individual pads, each composed of tens of lamellae which in turn incorporate hundreds of thousands of setae. Each seta is split into hundreds of spatula shafts ending with spatula pads at the sub-micrometer scale where the adhesive force is ultimately generated by van der Waals forces \cite{CW}.

Important insights into the functioning of microfibrillar adhesive devices in geckos were obtained from AFM (Atomic Force Microscopy) experiments at the scale of spatulas \cite{HUB} and setae \cite{AL,AD} and from attempts to artificially microfabricate fibrillar microstructures \cite{GD}.
 In particular, these studies revealed a strong dependence of the adhesive forces on the angle formed by the setal shafts with the adhesion surface, which suggests that easy detachment may be achieved by active reorientation of the single microscopic fibers. Theoretical understanding of the angle dependence of adhesion at the level of a single spatula is mostly based on the study of Kendall's model \cite{TP} which has been recently generalized  to account for  the asymmetric attachment-detachment behavior of a single seta \cite{GW} and  for  fiber tilting \cite{YD}. The main idea is that decohesion can be described as peeling which implies that a Griffith's fracture takes place in an infinitely localized tip of a steadily propagating crack. An alternative model suggested in \cite{AL,AD} links the cohesion-decohesion asymmetry with a dependence of the cohesive strength on the tangential component of the force. An interesting attempt to reconcile such friction-based approach with Kendall's fracture model was proposed in \cite{CWG} where a finite prestretch in the adhering layer was used to control the critical angle of adhesion.

While  fiber reorientation is clearly important for gecko adhesion, we propose in this paper a complementary mechanism  that can be broadly characterized as the possibility of  active switching between localized  (peeling) and delocalized  (pulling off) fracture. We argue that the organism can control the modality of detachment by changing the level of coupling  among individual fibrillar agents \cite{RU}.  In contrast to Kendall's model,  based on the assumption that the adhering layer can support only in-plane forces, we assume that this layer can have a shear stiffness which mimics bending resistance  and is responsible for the cooperative effects. We also assume that the organism can actively switch between the regimes of high and low stiffness depending on the force that has to be exerted. We discuss potential mechanisms of how such control at different scales of the fibrillar micro-structure can be achieved and propose a strategy of gecko advance. Based on the experimental scaling relations, we conjecture that the same adhesion mechanism is operative at every scale of the hierarchy and propose a simple model justifying the observed power law force-length relations.

\section{Discrete model}
\begin{figure}[h]
\includegraphics[height=1.7 cm]{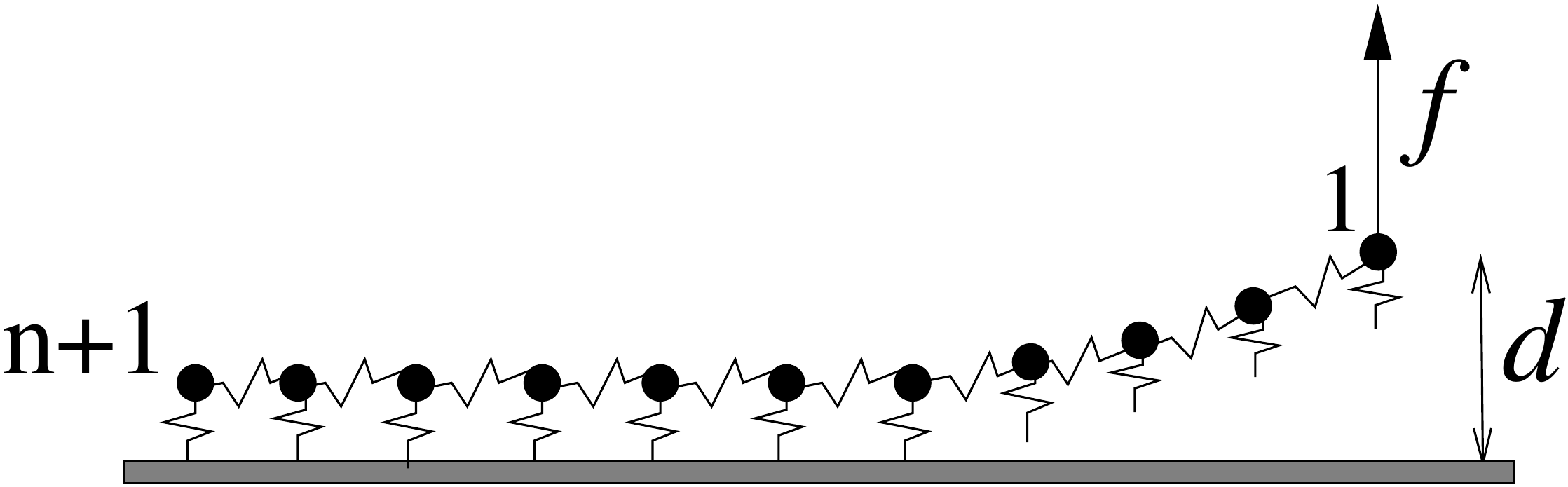}\vspace{0 cm}
\caption{\label{f1} Bishop-Peyrard model of decohesion.
}
\end{figure}

To describe an elemental fibrillar adhesion layer we use the minimal Bishop-Peyrard (BP) model \cite{Pey,Mc,MP}. Consider a chain with $n+1$  particles interacting through
$n$ linear springs and  bound to a rigid substrate by breakable elastic elements (see Fig.\ref{f1}).
Assume that the particles can move only in the vertical direction and denote by  $u_i$  the displacement of the particle
$i$. The coupling between individual adhesive devices is controlled by the elastic energy
 $$\phi_G(\delta_i)=\frac{1}{2} G \delta_i^2,$$
 where $G$ is the (shear)
stiffness, $\delta_i=(u_{i+1}-u_{i})/l$ is the strain and $l$ is the spring length. The cohesive energy at the micro scale can be modeled by a piece-wise quadratic function
 \be \phi_k(u_i )=\left \{ \ba{ll}
\frac{1}{2} \, k  u_i^2, &\mbox{if}\:\: \vk u_i< u_r, \\
  \gamma b,   &\mbox{if}\:\: u_i\geq u_r, \ea \right .
\label{ae}\ee \noindent where $k$ is the extensional stiffness, $\gamma$ is the adhesion energy density, $b$ is the out of plane spatial scale  and
 $ u_r=\sqrt{2\gamma b/k}\label{ur} $ is the limit displacement.

 We assume that the organism applies to the pad a localized loading.  Denote by $d$ the assigned displacement at  the  point $n=1$  and assume that all other points are unloaded. To find the  equilibrium force-displacement relation  $f(d)$ we minimize the energy
 \be  \Phi=l  (\sum_{i=1}^{n+1}  \phi_k(u_i)+ \sum_{i=1}^{n}
  \phi_G (\delta_i)).
  \ee
The corresponding energy landscape is complex \cite{MP}, however, we are interested only in  a set of local minimizers that can be parametrized by the position of the decohesion front $\xi\in [0,n],$
 \be
 \!u_i\!=\!\left \{
 \ba{ll} \displaystyle d - (i-1)l\, \frac{f}{G}, &  \!i=1,...,\xi, \vspace{0.2 cm}\\ \!\!
 \frac{\cosh [(n+3/2-i)\eta]l}{2\sinh [(n+1-\xi) \eta] \sinh
[\eta/2]}\displaystyle\, \frac{f}{G} , & \!i\!=\!\xi+1,...,n+1.\ea \right .
  \ee
Here the force is implicitly given by
\be f=\frac{2 n \nu^2}{(2\xi-1+\coth \frac{\eta}{2}\coth[(n+1-\xi)\eta])} \,\frac{d}{L}\, \, k\label{discr}\ee
where $\eta$
is a solution of $$1+l^2 /(2\nu^2)= \cosh[\eta]$$  and
$$\nu= \sqrt{G/k}$$ is the internal length scale, defining the size of the cohesive zone.
\begin{widetext}

  \begin{figure}[h]
\hspace{-0.3 cm}\includegraphics[width=12.0 cm]{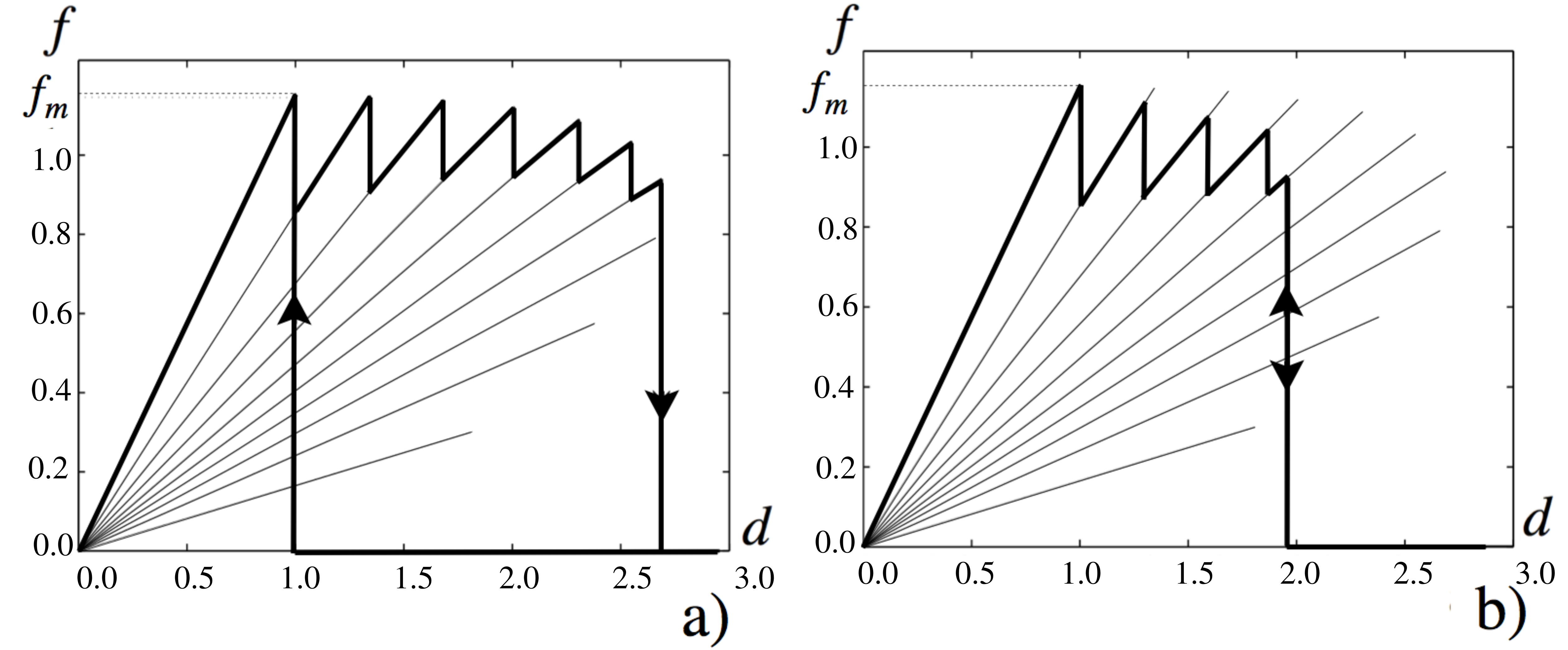}\vspace{-0.2 cm}
\caption{\label{f2b} Equilibrium force-displacement curves for a BP system with $n=10$ breakable links, $L=3$,
$u_r=1$, $k=1$, and $G=1$. Solid lines indicate metastable branches. Bold lines indicate force-displacement paths associated with the overdamped limit (a) and with global energy minimization (b). }\end{figure}
\end{widetext}

In Fig.\ref{f2b} we  show the metastable branches $f(d;\xi)$ parametrized by $\xi$; each branch ends at  $\bar d(\xi)$ satisfying   $u_{\xi+1}(\bar d(\xi);\xi)=u_r$. If the dynamics is overdamped  and  the driving  $d(t)$  is  quasistatic  we obtain the loading-unloading hysteresis indicated in Fig.\ref{f2b}(a) by bold lines. An alternative, hysteresis-free path of global energy minimization implied in Kendall's model is shown in Fig.\ref{f2b}(b). Observe that in both cases the cohesion force exhibits a characteristic plateau with which we can associate the maximal force $f_m$. The dependence of this threshold on the stiffness of the pad $G$ plays the main role in the proposed mechanism. The quasistatic assumption, which this model shares with Kendall's model, is supported  by experimental observations that the attachment/detachment rates are independent of the geckos speed \cite{AH}.
\begin{figure}[th]\vspace{-0.1 cm}
\hspace{-0.25 cm }\includegraphics[height=4.8 cm]{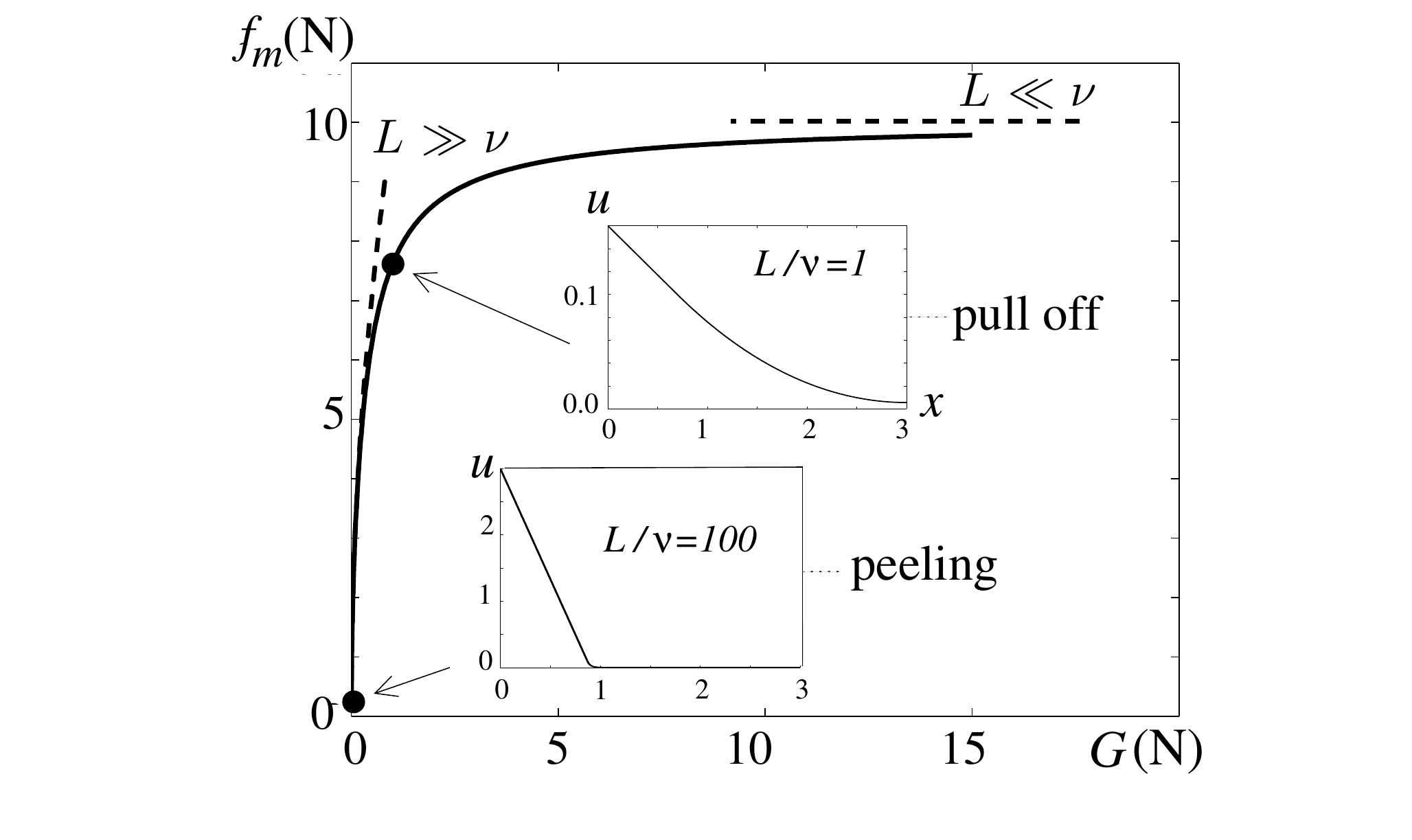}
\vspace{-0.5 cm}\caption{\label{f3}  Stiffness dependence of the adhesion threshold $f_{m}(G)$  for a system with $u_r=1$ mm, $k=1$ MPa and $L=10 $ mm. Insets:  crack configurations  illustrating localized ($L/\nu=100$) and diffuse  ($L/\nu=1$) cohesive zones.}
\end{figure}

\section{Continuum limit}
A simple analytical expression for the function $f_m(G)$ can be obtained in the continuum limit, which may be a poor approximation at the scale of the whole foot ($n\sim5$), but turns out to be fully adequate at the level of the setae ($n\sim50$). To study this limit we fix the total length $L=nl$ and assume that
 $n\rightarrow\infty$ and $l\rightarrow 0$. The continuum energy takes the form \be \Phi (u)= \int_0^L (\phi_k(u)+\phi_G(u'))\, dx \ee and its minimization is straightforward \cite{MT,MP}. Given the loading  $u(0)=d$ we obtain
\be
u_{ \zeta}(x)\!=\!\left \{ \ba{ll}
  d-\frac{x}{ \zeta L}(d-u_r) &
\mbox{ if } x \in (0, \zeta L ), \vks \\
\frac{ \cosh\frac{L-x}{\nu} }{\cosh \frac{L- \zeta L}{\nu}}\,\, u_r & \mbox{ if } x \in ( \zeta L , L) ,\ea \right .\ee
where $\zeta$ is the detached fraction of the pad which parametrizes the relation between the boundary displacement
$$ d=u_r+ \zeta L (f/G)$$ and the applied force
$$f=\sqrt{2\gamma\, b\, G}\tanh  [L(1-\zeta)/\nu]. $$
The maximum value of the force $f_m$, representing the
detachment threshold, is attained  at $ \zeta =1$ and  $d=u_r$:\be f_{m} (G)
=\sqrt{2\gamma\, b\, G} \tanh  (L\sqrt{k}/\sqrt{G}). \label{ft} \ee
In the limit when the external length scale is much larger than the internal length scale ($L\gg\nu$) we get the asymptotics
$$ f_m \sim \sqrt{2\gamma\, b\, G}. $$
 In the opposite case we obtain
$$
  f_m \sim \sqrt{2\gamma\, b\, k}L.
$$
The value of the critical force and the structure of the associated cohesive layers is shown in  Fig.\ref{f3}. Notice that for $L\gg\nu$ the crack has a narrow tip and this regime can be associated with the peeling mode of fracture. Instead for  $L\ll \nu$  the cohesive zone spreads along the whole pad and this regime can be associated with a pull off mode. Interestingly,  for $L\gg\nu$ the area under the hysteresis loop $$Q=2\gamma b L$$ representing the detachment energy is exactly twice as big as in the case when $L\ll \nu$; high dissipation at small force is due to much larger displacement. In this sense transition from pull off to peeling is similar to the transition from brittle to ductile fracture.

Our main assumption in what follows is that the stiffness of the linear springs, mimicking the stiffness of the gecko's pad, can be actively varied by the organism. The feasibility of such control is clear from the fact that the gecko rolls in for attachment (active shortening and thickening of the digits) and rolls out for detachment (active  lengthening and thinning of the digits).  It is also known that digital hyperextension anticipates each attaching and detaching event and that the musculo-tendinous system may influence single lamellae in the process of controlled rolling \cite{RU, AH}.  Experiments with  geckos \cite{RU},  frogs \cite{HB}, and ants \cite{F}  also show that at the macroscale the easy release is achieved through the localization of the cohesive region. Yet another argument in support of the active muscular control comes from the observation  \cite{RH}  that to simplify horizontal walking  geckos keep the hyperextended state and start activating the attachment mechanism only at sufficiently high slope  requiring stronger adhesion.

To get a rough estimate of the required stiffness variation  we observe that the adhesion force in geckoes is about one order of magnitude  less than the detachment threshold \cite{Dai}. According to (\ref{ft}) this corresponds to two orders of magnitude in stiffness variation, which is compatible with the data  on geckos forced to detach \cite{pugno}. Different physical mechanisms may be employed to regulate the coupling among adhesive fibrils at different structural levels. Thus, at the cellular level the dynamic filamentous actin network is known to be very soft at low stress,  but can stiffen up to three orders of magnitude in response to stresses \cite{Miz,Koe,Kaza}. Such stresses can be generated internally through molecular motors; moreover, constant remodeling allows the cytoskeleton to remain in a marginally stable state and easily switch between softening and stiffening \cite{Martens,Reichl}. At elevated stresses a quick transition to softening may also be related to the unfolding of the cross-linkers such as filamin \cite{Didonna}. Within muscle sarcomeres the effective stiffness can vary with the number of myosins attached to actin fibers and will also be affected by the unfolding-refolding transition in titin \cite{Linke}. Notice also that parameter $\nu$ controlling the mode of detachment can vary not only because of the stiffness $G$ but also because of the stiffness $k$. The latter depends on the aspect ratio of the adhesive elements  and may be controlled by the organism through capillarity induced self-assembly, modulated by secretion or evaporation of liquids responsible for the interaction between the fibrils \cite{Eisner,Pokroy}.

At the macroscale, the variability of shear modulus may be associated with the reversible development of micro defects inside the tissue architecture with subsequent internal healing of this damage through remodeling. Following a classical approach in damage mechanics \cite{DD,DM}, we can introduce an internal variable $\alpha$, with $\alpha=0$ representing the undamaged state (stiffest configuration) and $\alpha=1$ -- the damage-saturated state (most compliant configuration). Then $G=\hat G(\alpha)$ with $\hat G'(\alpha)<0$. Since the detachment force depends on the level of damage $$\hat f_m(\alpha)=f_m(\hat G(\alpha)),$$ while the critical displacement does not ($d_m=u_r$), we can write the effective elastic stiffness for the attached state in the form $$\hat E(\alpha)=\sqrt{k\hat G(\alpha)}\tanh [L \sqrt{k}/\hat G(\alpha)].$$ Observe that $\hat E'(\alpha)<0$.
\begin{figure}[th]\vspace{-0.4 cm}
\includegraphics[height=6.2 cm]{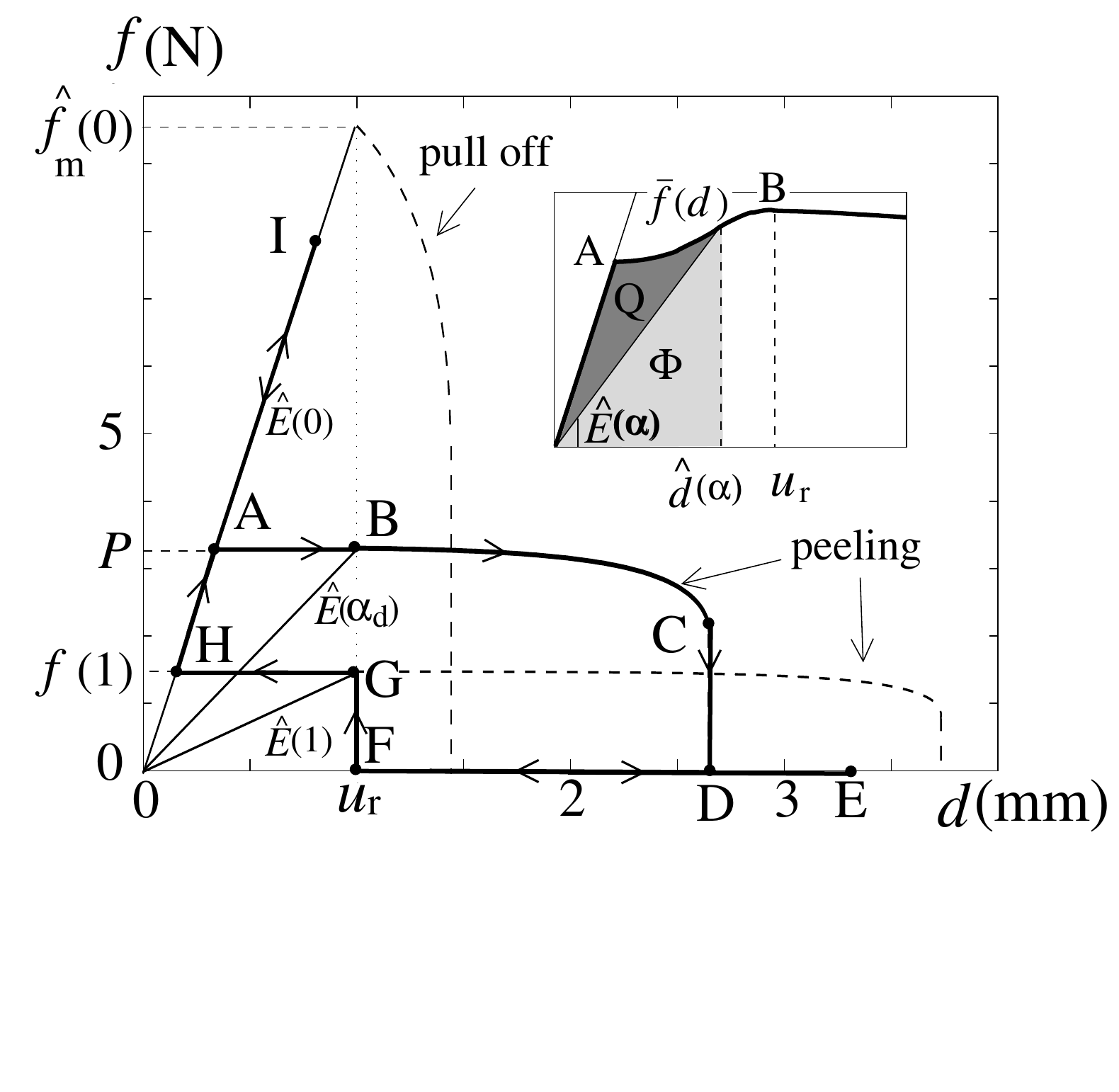}\vspace{-1.25 cm} \caption{\label{f4}  Schematic representation of the proposed attachment-detachment strategy (see text). The inset shows a generic force-displacement path  associated with stiffness variation AB and the decomposition of the corresponding external work $W$ into the elastic energy $\Phi$ and the dissipated energy $Q$.}
\end{figure}

\section{Attachmnt-detachment cycle}
We can now propose a mechanical strategy of geckos advance. Assume that only one scale of the fibrillar microstructure is involved, that the response is quasistatic and overdamped, and that the continuum limit is valid. To fix ideas, consider ceiling walking and suppose that the cycle of attachment-detachment  starts at an attached state A (see inset in Fig. \ref{f4}) where the pad is in the stiff configuration $\hat G(0)$ associated with the high threshold $\hat f_m(0)$. To attain the detachment below this threshold, gecko can induce a reversible ``damage process'' described by a force-displacement relation $f=\bar f(d)$ where $d=\hat d (\alpha)$ and $\hat E(\alpha)=\bar{f}(\hat d(\alpha))/\hat d(\alpha)$. In the process of stiffness variation the external work $$W=\int_{P/\hat E(0)}^{\hat d(\alpha)} \bar f(\tilde{d}) \, \, d\, \tilde{d}>0$$ (done by gravity) is partially stored as elastic energy $$\ljump \Phi\rjump=P^2/(2\hat E(0))- \hat E^2(\alpha)\hat d(\alpha)/2$$ and the rest is dissipated into heat $$Q=-\int_0^{\alpha} \frac{1}{2}\hat E'(\tilde \alpha)\hat d^{\, 2}(\tilde{\alpha})  d\tilde{\alpha}.$$ To minimize dissipation while maintaining a stable mechanical response with $ \bar f'(d)\geq 0$, the animal must ensure that damage advances at constant force $\bar f(d)=P$ (AB in Fig.\ref{f4}). In this case exactly half of the work is dissipated and we can write  $$\ljump \Phi\rjump=Q=(P/2)(u_r-P/\hat E(0))>0.$$ Damage at constant stress has been observed in many rubber-like materials and linked to inherent energy non-convexity \cite{DPS};  reversible structural changes at constant stress are also characteristic for muscle tetanus \cite{GHJ}.

After the critical force is decreased, the pad can be pulled away by peeling (path BC in Fig.\ref{f4}).  The detachment ends with an abrupt decohesion at point C in Fig.\ref{f4}. In order to reattach, the gecko can follow a reverse path EF.
As the foot is placed on the surface, the displacement $d$ is gradually decreased and the attachment takes place at the point F through ``inverse peeling'' at $d=u_r$. An instantaneous force jump brings the system to the point G and allows the animal to place some weight on the foot. To secure a robust attachment, the gecko must reverse the damage and induce  active stiffening (trajectory GH).

The system heals the damage by remodeling the damaged  configuration with $\alpha=1$ back into the virgin configuration with $\alpha=0$ while increasing stiffness and decreasing displacement. This requires work which is now done  by  the gecko. The energy comes from metabolic sources $M <0$ and is released due to elastic unloading  $$\ljump \Phi \rjump = (\hat f_m(1)/2)(\hat f_m(1)/\hat E(0)-u_r)<0.$$ If we again assume that healing (remodeling) takes place at constant stress (minimum metabolic energy path), we obtain $\ljump \Phi \rjump=M$. Experiments show that geckos in the compliant state have a very low detaching threshold $\hat f_m(1)$  \cite{Autlib} which means that the   metabolic energy required for such stiffness increase is also low.

After the state of high stiffness  is reached (point H in Fig.\ref{f4}) the peeling mode is deactivated because the  force required for detachment $\hat f_m(0)$ is now large. Therefore more weight can be shifted to this foot (path HI) and another foot can undergo the detaching-attaching cycle. Overall, the  detachment process BCD with decreasing force on the two detaching feet must take place simultaneously with the attachment process FGHA involving the other two reattaching feet  \cite{AH}.

\begin{figure}[th]\vspace{-0.15 cm}
\includegraphics[height=9.5 cm]{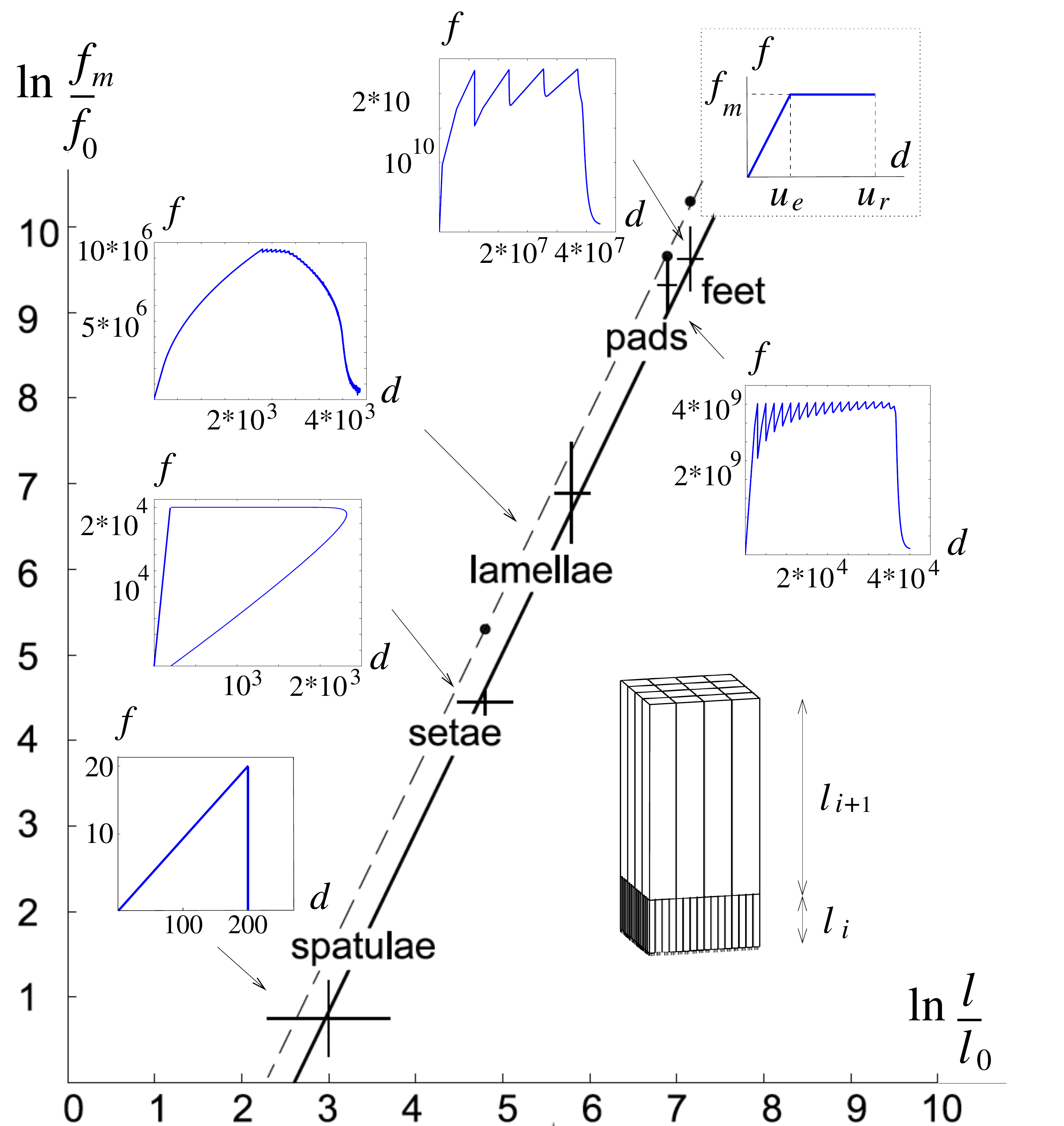} \vspace{-0.15 cm}
\caption{\label{f5} Scale free nature of the adhesion mechanism. Solid line interpolates the experimental values of the adhesion forces \cite{IL, CW, pugno, HUB, GW, AD, Autlib, AL, SN}; dashed line, of the friction forces \cite{Autlib}. Insets show nested  computations based on the BP model with adhesion links behaving as shown in the right upper corner. At the smallest scale of the elemental fibrilla  $u_r=200$ nm, $L=10^3$ nm and $k=100 $MPa.  The parameters $u_e$, $u_r$, and $k$   at larger scales have been computed iteratively. Other parameters: fibrilla $\rightarrow$ seta, $n=50$, $L=100$ nm, $G=0.2$ mN; seta $\rightarrow$ lamella, $n=50, L=0.5$ mm, $G= 1$ N, lamella $\rightarrow$ pad, $n=20, L=3$ mm, $G=300$ N, pad $\rightarrow$ foot: $n=5, L=20$ mm, $G=20$ N. } \end{figure}

\section{Hierarchical architecture}
To understand the role played by the hierarchical  micro-fibrillar architecture of the adhering pad  \cite{GW, KB} we compare adhesion forces at different spatial scales (see Fig.\ref{f5}). Using the experimental data \cite{IL, CW, pugno, HUB, GW, AD, Autlib, AL, SN} we may deduce a scaling relation $$ f_{m}\sim l^\beta,$$
with exponent $\beta=2\pm 0.5$. In Fig.\ref{f5} we show the power law together with the scaling of friction forces reported in \cite{Autlib}; our results are consistent with the observation  that the friction force is usually five times higher than the adhesion force \cite{pugno}.

While the exact origin of these empirical  power law relations is not known, we can propose the following plausible explanation. Following \cite{JB} we suppose that the architecture of the fibrillar adhesive system is designed to ensure that in contact with rough surfaces the fibrils buckle simultaneously at all scales to ensure maximum folding which accommodates fractal roughness.  If at hierarchical level $i$, with spatial scale $l_i$, the system is marginally stable against buckling, then  $f_i =c E l_i^2$, where $E$ is the elastic modulus and the constant $c$ depends on the shape of the cross section, aspect ratio, and the boundary constraints. We assume the simplest allometric law when both $c$ and $E$ are scale invariant. Now, if the fibrils at a finer level $i$  cover the tips of the fibrils at the coarser level $i+1$ densely, which is known as Leonardo's rule \cite{Eloy}, then $l_{i+1}^2=n_i l_{i}^2$ where $n_i$ is the number of fibrils at the level $i$ (see the scheme at the bottom right corner of Fig.~\ref{f5}). Since $f_{i+1}=n_i f_{i}$, we obtain $f_{i+1}=c E l_{i+1}^2$, which means that the force is critical also at the $i+1$ level.  In view of our neglect of collective modes of instability \cite{R}, we can only tentatively conclude from this reasoning that the observed scaling supports the idea that the whole structure can be marginalized simultaneously. An important feature of this scaling, however, is that stress is uniform which has been proposed previously as a criterion of optimality in several biological and engineering  systems \cite {A}.

The power law scaling is indicative of a scale-free detachment mechanism.  We can model it in our framework by the appropriate renormalization of the parameters in a scale-generic BP model shown in the upper  corner of Fig.\ref{f5}, where $u_e$ is the elastic threshold  and $u_r$ is the  detachment displacement. The parameters $u_e$, $u_r$, and $k$ can  be computed at each scale iteratively by using a series of nested BP models,  whereas the parameter $G$ characterizing the elastic coupling  can be chosen at each scale to match the experimentally measured detachment threshold.

At the smallest scale of a spatula  the adhesive properties of the fibrils can be described  by the energy density (\ref{ae}) with parameters $u_r$ and $k$ available from  experiment \cite{HUB}. The behavior at the next scale (setae) can be obtained numerically  by simulating an overdamped gradient flow dynamics for a quasistatically driven BP system with $n=50$ spatulae;  the value of $G$ is chosen to ensure the maximum measured adhesion force of $40 {\rm \mu N}$ \cite{AL}. The overall behavior at this scale matches the experimental results in \cite{AL} showing an elastoplastic range ending with an abrupt detachment. At the level of a  lamella, we consider $n=50$ elastic-plastic elements with constitutive parameters obtained from the spatulae scale simulations. At the next level of a toe, we need to model lamellae with $n=20$ and finally at the level of a foot $n=5$ (toes) and the problem becomes strongly discrete. The results at this last level match observations showing digitized  detachment of the toes \cite{RU}.

\section{Conclusions} By studying a prototypical system, we have shown that both the force threshold and the dissipation associated with reversible adhesion can be modified by active control of the coupling among individual adhering elements.   The possibility of the ensuing multi-path adhesion \cite{B} reconciles strong binding with easy debinding, which is at the base of the observed agility of  lizards and insects running on inclined surfaces.  The proposed mechanism has a scale-free hierarchical structure, which is typical for biological systems at all levels of organization.

\section*{Acnowledgments} The authors thank F. Maddalena and D. Percivale for helpful 
discussions. G.P. work has been supported  by PRIN 2010-11: ``Dinamica, 
stabilit\`a e controllo di strutture flessibili''.

\bibliographystyle{apalike}
{\fontsize{10}{10}\selectfont

\end{document}